# 'Antigravity' propulsion and relativistic hyperdrive

Franklin S. Felber[a)]

*Physics Division, Starmark, Inc., P. O. Box 270710, San Diego, California 92198*

Exact payload trajectories in the strong gravitational fields of compact masses moving with constant relativistic velocities are calculated. The strong field of a suitable driver mass at relativistic speeds can quickly propel a heavy payload from rest to a speed significantly faster than the driver, a condition called hyperdrive. Hyperdrive thresholds and maxima are calculated as functions of driver mass and velocity.

This paper analyzes the exact unbounded orbits of a payload in the gravitational field of the much greater mass of a source moving at constant velocity. The gravitational field of a suitable compact source can accelerate a payload from rest to relativistic speeds faster than that of the source.

A recent paper [1] showed that at relativistic speeds, a suitable driver mass can quickly propel a heavy payload from rest nearly to the speed of light with negligible stresses on the payload. That paper analyzed the payload motion only for drivers and payloads approaching each other directly, that is, with zero impact parameter and zero angular momentum in the system. This paper analyzes the exact strong-field orbital dynamics of a payload in the potentially more useful case of a close encounter, but not a collision, of a relativistic driver with an initially stationary payload.

In this paper, as in [1], the difficulties of calculating the exact strong gravitational field of a relativistic mass are by-passed completely. Instead, the propulsion of a payload is calculated directly by a two-step procedure, without calculating the field that produced the motion. The first step is to calculate the well-known exact trajectory of a payload in the static Schwarzschild field of a stationary, spherically symmetric mass. The second step is to perform a simple Lorentz transformation from the rest frame of the driver mass to the initial rest frame of the payload when the payload is far from the driver.

This two-step procedure for calculating the propulsion of a payload from rest allows no possibility for confusion or ambiguity of coordinates, because it focuses on the observed physical payload trajectory, rather than the field that produced it. The Schwarzschild solution is a unique spherically symmetric solution of Einstein's equation for which the metric becomes asymptotically flat as the distance from the source increases to infinity. The spacetime coordinates of the Schwarzschild solution, $r$, $\theta$, $\phi$, and $t$, are the coordinates used by a *distant inertial observer* to observe and measure the trajectory of a payload in a Schwarzschild field. A *distant inertial observer* is defined as an unaccelerated observer in flat spacetime far beyond any gravitational influence. *Every pair of distant inertial observers of a payload trajectory must agree on the position vs. time of the payload, differing in their accounts only by a simple Lorentz transformation.* In particular, the payload trajectory observed by a *distant inertial observer* in the rest frame of the driver mass must differ only by a simple Lorentz transformation from the same trajectory observed by a *distant inertial observer* in the initial rest frame of the payload (in which the payload is still beyond the gravitational influence of the driver).

Although this paper does not calculate the strong field of a relativistic mass, a recent calculation [2] of the exact *weak* field of a mass in arbitrary relativistic motion suggested this means of accelerating massive payloads close to the speed of light. This new 'antigravity' solution [2] of Einstein's gravitational-field equation was the first to calculate the changing gravitational field of masses moving near the speed of light, $c$. The solution showed that any mass faster than $3^{-1/2}c$, no matter how light or how distant, will be seen by a *distant inertial observer* to repel other masses lying within a narrow 'antigravity beam.' The closer a mass gets to the speed of light, the stronger its 'antigravity beam' becomes.

This new means of 'antigravity' propulsion addresses the major engineering challenges for near-light-speed space travel of providing enormous propulsion energy quickly without undue stresses on a spacecraft. By conventional propulsion, acceleration of a 1-ton payload to $0.9c$ requires imparting a kinetic energy equivalent to about 30 billion tons of TNT. In the 'antigravity beam' of a speeding star or compact object, however, a payload would draw its energy for propulsion from the repulsive force of the much more massive driver. Moreover, since it would be moving along a geodesic, a payload would 'float weightlessly' in the 'antigravity beam' even as it was accelerated close to the speed of light.

As seen by a *distant inertial observer*, the gravitational field of a driver mass, moving faster than $3^{-1/2}c$, repels other masses lying within a narrow cone angle. This threshold velocity for an 'antigravity' field was one outcome of the first solution of Einstein's gravitational-field equation for a time-dependent relativistic source [2]. Liénard-Wiechert retarded-potential methods in the weak-field approximation were used to calculate the gravitational field at a stationary point in spacetime of a mass in arbitrary relativistic motion. In this approximation, the field at a stationary spacetime point is the same as the field on a payload moving freely along a geodesic, as long as the payload starts from rest. That is, the 'gravitomagnetic' terms in the geodesic equation are of the same order as terms that are neglected in the weak-field approximation.

Earlier solutions of Einstein's equation for the gravitational field of moving sources applied either to relativistic or time-dependent sources, but not both. For example, the Kerr solution [3–7] applies to a relativistic rotating axisymmetric source, but the Kerr field is stationary. Earlier calculations of the gravitational fields of arbitrarily moving masses were done only to first order in the ratio of source velocity to the speed of light [5,7–10]. Even in a weak static field, these earlier calculations have only solved the geodesic equation for a *nonrelativistic* test particle in the *slow-velocity* limit of source motion. In this slow-velocity limit, the field at a moving test particle has terms that look like the Lorentz field of electromagnetism, called the 'gravimagnetic' or 'gravitomagnetic' field [7,9,10]. Harris [9] derived the *nonrelativistic*

---

[a)]Electronic mail: starmark@san.rr.com

equations of motion of a moving test particle in a dynamic field, but only the dynamic field of a *slow-velocity* source. Only when Einstein's equation is solved for relativistic, time-dependent sources, as in [1] and [2], can even weak gravitational fields be shown to be repulsive.

The threshold velocity of $3^{-1/2}c$ for an 'antigravity' field can be derived in a number of ways, including the following:

(1) *Equation of motion in Schwarzschild field.* The spacetime interval for the spherically symmetric Schwarzschild field of a mass $M$ is $ds^2 = \psi dt^2 - \psi^{-1} dr^2 - r^2 d\theta^2 - r^2 \sin^2\theta d\phi^2$, where $\psi(r) \equiv 1 - r_s/r$ is the $g_{00}$ component of the Schwarzschild metric, and $r_s \equiv 2GM/c^2$ is the Schwarzschild radius. From this interval, the equations of motion of a payload in coordinate time $t$ are found to be [1,7]

$$\beta_r^2 = \psi^2 - \left(1 + L^2/c^2 r^2\right)\psi^3/\gamma_0^2, \quad (1)$$

$$\beta_\phi = L\psi/\gamma_0 cr, \quad (2)$$

where $\beta_r \equiv \dot{r}/c$ and $\beta_\phi \equiv r\dot{\phi}/c$ are the radial and azimuthal components of the normalized payload velocity $\boldsymbol{\beta}(r)$ measured by a *distant inertial observer*; an overdot indicates a derivative with respect to $t$; $\gamma_0 c^2$ is the conserved total specific energy of the payload; and $L$ is its conserved specific angular momentum. If the payload has a speed $\beta_0 c$ far from the mass $M$, where $\psi \approx 1$ and $\dot{r} \gg r\dot{\phi}$, then from Eq. (1) we find $\gamma_0^2 = 1/(1-\beta_0^2)$. In this same weak-field limit of large $r$, a virtual radial displacement $\delta r$ results in a change of $\beta_r^2$ by an amount

$$\delta \beta_r^2 \to (3\beta_0^2 - 1)(r_s/r^2)\delta r, \quad (3)$$

indicating a gravitational repulsion when $\beta_0 > 3^{-1/2}$, whether the payload is displaced towards or away from the source $M$.

The right-hand-side of Eq. (1) is a normalized conservative potential, from which is derived the exact equation of motion in the Schwarzschild field [1],

$$\ddot{r} = \frac{-GM}{r^2}\left(\frac{3\psi^2}{\gamma_0^2} - 2\psi\right) + \frac{L^2 \psi^2}{\gamma_0^2 r^3}\left(\psi - \frac{3r_s}{2r}\right). \quad (4)$$

For purely radial motion of a payload ($L = 0$), the gravitational force is repulsive when $\beta_0^2 > 1 - 2/3\psi$. From this inequality, we see that even a weak Schwarzschild field is always seen by a *distant inertial observer* to repel a payload moving radially whenever $\beta_0 > 3^{-1/2}$, and a strong field is seen to repel a payload whenever it is within $3r_s$ of a stationary compact object, even at slow velocities. The repulsive force measured by a *distant inertial observer* is the same whether the payload moves radially inwards or outwards, because the acceleration in Eq. (4) depends only on the square of $\beta_0$, and not on its sign.

(2) *Liénard-Wiechert retarded field.* In the weak-field approximation, a general expression was derived in 3-vector notation for the retarded gravitational field $\mathbf{g}(\mathbf{r},t)$ of a source in arbitrary relativistic motion [2]. From this expression, the retarded field of a source moving with constant velocity $c\boldsymbol{\beta}_0$ is given in the forward (upper sign) and backward (lower sign) directions by

$$g_\pm(\mathbf{r},t) = -\gamma_0^5 (1 \pm \beta_0)^2 (1 - 3\beta_0^2) GM/\{R^2\}_{ret}, \quad (5)$$

where $\{R\}_{ret}$ is the separation of the source and the observation point $(\mathbf{r},t)$ at the retarded time $t - \{R\}_{ret}/c$. Although the retarded field is greater in the forward direction than the backward direction by a factor $(1+\beta_0)^4 \gamma_0^4$, the field is repulsive in both directions when $\beta_0 > 3^{-1/2}$.

(3) *Geodesic equation with dynamic metric.* The weak-field dynamic (time-dependent) metric on the $x$ axis of a source moving with constant velocity $c\boldsymbol{\beta}_0$ along the $x$ axis was derived from Einstein's equation in the harmonic gauge [1]. The spacetime interval on the $x$ axis in Cartesian coordinates is [1]

$$ds^2 = [1 + 2(1+\beta_0^2)\Phi]dt^2 - [1 - 2(1+\beta_0^2)\Phi]dx^2 - 8\beta_0 \Phi dxdt - [1 - 2(1-\beta_0^2)\Phi](dy^2 + dz^2), \quad (6)$$

where the dimensionless potential, $\Phi \equiv -GM\gamma_0/(x-\beta_0 ct)c^2$, satisfies the harmonic gauge condition, $\partial \Phi/\partial t + c\beta_0 \partial \Phi/\partial x = 0$. In the weak-field approximation, in which terms of order $\Phi^2$ are neglected, the geodesic equation gives the acceleration of a payload as [1]

$$d^2x/dt^2 \approx -\gamma_0 (1 - 3\beta_0^2) GM/(x - \beta_0 ct)^2, \quad (7)$$

which is repulsive in both directions, $+x$ and $-x$, for $\beta_0 > 3^{-1/2}$. Since the retarded-time separation of driver and payload is $\{R\}_{ret} = (x - \beta_0 ct)/(1 \mp \beta_0)$, Eq. (7) is identical to Eq. (5). The equivalence of Eqs. (5) and (7) demonstrates the correspondence of the Liénard-Wiechert weak-field solution for sources in arbitrary motion in [1] with the direct solution of Einstein's equation and the geodesic equation for sources in uniform motion in [2].

(4) *Lorentz transformation of trajectory in Schwarzschild field.* Animated solutions accompanying [1] graphically illustrate the 'antigravity' threshold for purely radial motion at $\beta_0 = 3^{-1/2}$, even in weak fields, found by the two-step approach presented in [1].

Figure 1 illustrates the two-step approach of this paper and of [1] to calculating the exact unbounded orbital motion of a payload mass $m$ in the field of a driver mass $M$ with constant velocity $c\boldsymbol{\beta}_0$. First, the unique trajectory in the static Schwarzschild field of a stationary mass $M$ is found for which the perigee, the distance of closest approach, of the payload is $b$ and the asymptotic velocity far from the stationary mass is $-c\boldsymbol{\beta}_b$. In the static field of $M$, the trajectory is time-reversible. Second, the trajectory is Lorentz-transformed to a reference frame moving with constant velocity $-c\boldsymbol{\beta}_0$, in which the mass $M$ has a constant velocity $c\boldsymbol{\beta}_0$, and the payload far from the driver is *initially* at rest. The Lorentz trans-formation occurs between two *distant inertial observers* far from the interaction, in asymptotically flat spacetime.

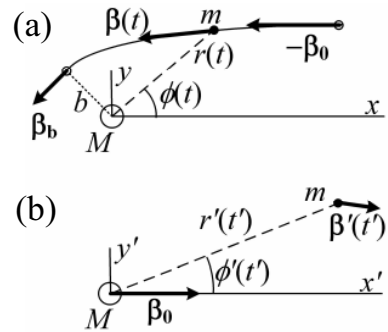

FIG. 1. Two-step exact solution of 'antigravity' propulsion of payload mass $m$ by relativistic driver mass $M$: (a) Schwarzschild solution in static field of $M$. (b) Solution Lorentz-transformed to initial rest frame of payload far from $M$. Primes denote Lorentz-transformed quantities.

In the dimensionless coordinate, $\rho \equiv b/r$, the orbital equation of the payload in the Schwarzschild field is [1,6,7]

$$\left(\frac{d\rho}{d\phi}\right)^2 = (1-\alpha)\left(\frac{\gamma_0^2 - 1 + \alpha\rho}{\gamma_0^2 - 1 + \alpha}\right) - (1-\alpha\rho)\rho^2, \quad (8)$$

where $\alpha \equiv r_s/b$, ranging from 0 to 1, is a measure of the field strength at perigee.

Integrating Eq. (8) gives the exact payload trajectories shown in Fig. 2(a). The Cartesian coordinates, $x$ and $y$, in Fig. 2(a) are normalized to $r_s$. In Fig. 2(a), the three payload trajectories were taken to begin far to the right of the figure (at $x \gg 10$) with the same velocities, $\beta_0 = 0.9$ in the $-x$ direction, but with different impact parameters that resulted in total deflections of $30°$, $90°$, and $180°$.

Figure 2(b) shows the same three trajectories after a Lorentz transformation by $0.9c$ to the initial rest frame of the payloads, in which the driver mass has constant velocity $\beta_0 = 0.9$ in the $+x$ direction. The animated versions of Figs. 2(a) and 2(b), available at [11], clock the payload speeds along each trajectory seen by *distant inertial observers* in both reference frames. In the animations, all payloads start from $x = 10$ at $t = 0$ and from $x' = 10/\gamma_0$ at $t' = 0$.

Since trajectories are time reversible in a Schwarzschild field, the final payload speed equals the initial speed, $\beta_0$. A total deflection of $180°$ results in a complete U-turn and final payload velocity of $\beta_0$ in the $+x$ direction. When Lorentz-transformed, as in Fig. 2(b), this U-turn trajectory results in the maximum final hyperdrive velocity, $\beta'_{max} = 2\beta_0/(1+\beta_0^2)$, in the $+x'$ direction. As shown in Fig. 3, this maximum hyperdrive velocity imparted to the payload corresponds to a relativistic factor, $\gamma'_{max} = (1+\beta_0^2)\gamma_0^2$, much greater than the relativistic factor of the driver mass, $\gamma_0$, if the driver speed is close to the speed of light.

The threshold field strength for hyperdrive is taken to be the value of $\alpha$ that results in propulsion of a payload from rest to the same speed, $\beta' = \beta_0$, as the driver mass. Figure 4 shows the field-strength parameter $\alpha$ needed to impart to an initially stationary payload the threshold hyperdrive speed, $\beta' = \beta_0$, and the maximum hyperdrive speed, $\beta'_{max} = 2\beta_0/(1+\beta_0^2)$, as a function of $\beta_0$.

Integrating Eq. (8) in the nonrelativistic Newtonian limit, in which $\beta_0^2 \ll 1$ and $\alpha \ll 1$, gives the total deflection, $\Delta\phi$, of an unbounded payload orbit in a Schwarzschild field as $\sin(\Delta\phi/2) = (1+2\beta_0^2/\alpha)^{-1}$. In this limit, the final speed that is imparted to an initially stationary payload by a moving driver is $\beta' = 2\beta_0/(1+2\beta_0^2/\alpha)$. In this nonrelativistic Newtonian limit, therefore, the maximum final speed that can be imparted in the initial rest frame of the payload is $\beta'_{max} = \beta_0$, which occurs for $\beta_0 = \sqrt{\alpha/2}$ and $\Delta\phi = 60°$.

Integrating Eq. (8) in the limit $\alpha \ll \beta_0^2$, which allows relativistic velocities, gives the total deflection of an unbounded payload orbit in a weak Schwarzschild field as $\Delta\phi = \alpha(1+1/\beta_0^2)$, which corresponds to the deflection of a photon for $\beta_0 = 1$. In the limit of highly relativistic driver speeds ($\gamma_0 \gg 1$), even a weak field can accelerate a payload from rest to the same speed as the driver. In this limit, the threshold field strength needed for hyperdrive is $\alpha = (2\gamma_0)^{-1/2}$. As shown in Fig. 4, however, a strong field of $\alpha > 0.5681$ is needed for a highly relativistic driver to accelerate a payload from rest to the maximum hyperdrive speed.

In conclusion, a simple Lorentz transformation of the well-known unbounded orbit of a payload in a Schwarzschild field gives the exact payload trajectory in the strong field of a relativistic driver with constant velocity, as seen by a *distant iner-*

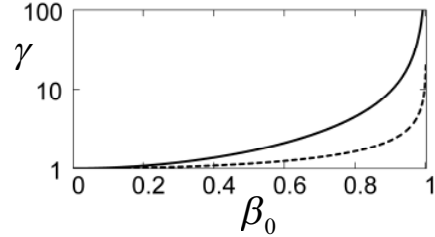

FIG. 3. Payload specific energy $\gamma'_{max}$ after maximum hyperdrive (solid) and driver specific energy $\gamma_0$ (dashed) vs. driver speed $\beta_0$.

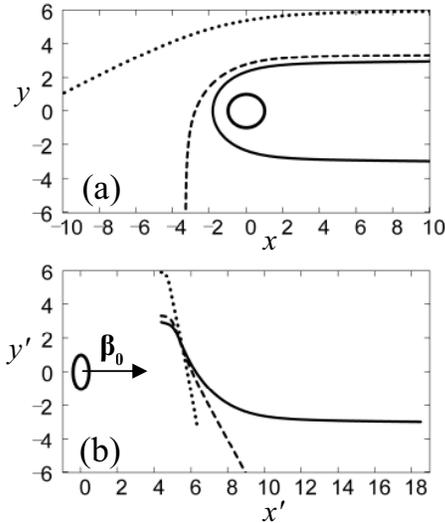

FIG. 2. Unbounded ballistic trajectories: (a) In strong Schwarzschild field with initial velocities $\beta_0 = 0.9$ in $-x$ direction and total deflections $30°$ (dotted), $90°$ (dashed), and $180°$ (solid). Scale normalized to $r_s$. (b) After Lorentz transformation of same trajectories to initial rest frame of payloads. Both (a) and (b) are animated in [11].

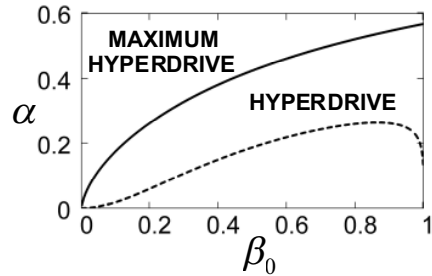

FIG. 4. Field-strength parameter $\alpha$ required to accelerate payload from rest to maximum hyperdrive speed (solid) and to hyperdrive threshold $\beta_0$ (dashed) vs. driver speed $\beta_0$.

*tial observer*. The calculations of these payload trajectories by this two-step approach, and their animated versions [11], clearly show that suitable drivers at relativistic speeds can quickly propel a heavy payload from rest to speeds close to the speed of light.

The strong field of a compact driver mass can even propel a payload from rest to speeds faster than the driver itself – a condition called hyperdrive. Hyperdrive is analogous to the elastic collision of a heavy mass with a much lighter, initially stationary mass, from which the lighter mass rebounds with about twice the speed of the heavy mass. Hyperdrive thresholds and maxima were calculated and shown in Fig. 4 as functions of driver mass and velocity. Substantial payload propulsion can be achieved in weak driver fields, especially at relativistic speeds. However, the maximum relativistic factor, $\gamma'_{max} = (1 + \beta_0^2)\gamma_0^2$, that a nonrotating driver can impart by hyperdrive to an initially stationary payload can only be produced at relativistic speeds by the strong gravitational field of a compact driver mass, as shown in Figs. 3 and 4.

Although the calculation is beyond the scope of this paper, we expect that even higher payload speeds can be achieved in the strong field of a spinning compact mass. Inertial frame dragging in the strong Kerr field of a rotating compact mass [3–7] should impart angular momentum and energy to a payload orbiting in the same direction as the rotation.